\documentclass[preprint,12pt]{elsarticle}

\usepackage{amssymb}

\usepackage[colorlinks=True,urlcolor=blue,linkcolor=blue,citecolor=blue]{hyperref}

\journal{Infared Physics \& Technology}

\begin{document}

\begin{frontmatter}



\title{A broadband silicon quarter-wave retarder for far-infrared spectroscopic circular dichroism}


\author[bnl]{Xiaoxiang~Xi\corref{cor1}}
\ead{xiaoxiang@bnl.gov}
\author[bnl]{R.J.~Smith}
\author[njit]{T.N.~Stanislavchuk}
\author[njit]{A.A.~Sirenko}
\author[vu]{S.N.~Gilbert}
\author[ccny]{J.J.~Tu}
\author[bnl]{G.L.~Carr}

\cortext[cor1]{Corresponding author. Tel.: +16313448727; Fax: +16313447039;}
\address[bnl]{Photon Sciences, Brookhaven National Laboratory, Upton, NY 11973, USA}
\address[njit]{Department of Physics, New Jersey Institute of Technology, Newark, NJ 07102, USA}
\address[vu]{Department of Physics, Vanderbilt University, Nashville, TN 37235, USA}
\address[ccny]{Department of Physics, the City College of the City University of New York, New York, NY 10031, USA}

\begin{abstract}
The high brightness, broad spectral coverage and pulsed characteristics of infrared synchrotron radiation enable time-resolved spectroscopy under throughput-limited optical systems, as can occur with the high-field magnet cryostat systems used to study electron dynamics and cyclotron resonance by far-infrared techniques. A natural extension for magnetospectroscopy is to sense circular dichroism, i.e. the difference in a material's optical response for left and right circularly polarized light.  A key component for spectroscopic circular dichroism is an achromatic $^1/_4$ wave retarder functioning over the spectral range of interest.  We report here the development of an in-line retarder using total internal reflection in high-resistivity silicon. We demonstrate its performance by distinguishing electronic excitations of different handednesses for GaAs in a magnetic field. This $^1/_4$ wave retarder is expected to be useful for far-infrared spectroscopy of circular dichroism in many materials.
\end{abstract}


\end{frontmatter}


\section{Introduction}
\label{intro}
Infrared spectroscopy is a well-developed technique to study electronic and vibrational properties in a variety of material systems of interest in condensed matter physics, chemistry, and biology. The energy levels associated with electronic transitions or vibrational modes manifest as spectroscopic features at photon energies specific to each material, enabling material identification and characterization. The capabilities of infrared spectroscopy can be greatly enhanced by using polarized light to reveal further details of the excitations. For example, linearly polarized infrared light has been commonly employed to distinguish lattice vibrational modes and electronic excitations along different directions in basic anisotropic crystals.  

When the excitation involves the orbital motion of electrons in a magnetic field, or their spin, the absorption of light is usually limited to circularly polarized light of a particular handedness.  The difference in the resulting optical response between left and right circularly polarized light is referred to as circular dichroism.  The effect can occur for various novel quantum states in condensed matter physics and material science. For example, Landau level spectroscopy in monolayer \cite{
Witowski2010,Pound2012,Morimoto2012,Falkovsky2013}, bilayer \cite{Pound2012,Morimoto2012,Falkovsky2013,Abergel2007,Ho2010,Bisti2011}, and multilayer \cite{Sadowski2006,Koshino2008,Crassee2011} graphene, graphite \cite{Chuang2009,Orlita2009}, as well as giant Rashba spin-splitting semiconductors \cite{Bordacs2013} can probe the electronic energy levels quantized by a magnetic field. Circularly polarized light can distinguish transitions with opposite handedness, putting strict constraints on the selection rules for the optical transitions. Faraday and Kerr rotations, caused by different propagation velocities for left and right circularly polarized light in a medium, have been reported in graphene \cite{Shimano2013}. More intriguingly they have been proposed to experimentally verify the topological magnetoelectric effect in 3D topological insulators, a phenomenon of topological quantization in the unit of the fine structure constant \cite{Maciejko2010}. Availability of left and right circularly polarized light in principle allows the measurement of the off-diagonal optical conductivity, the quantity underlying the Faraday and Kerr rotations. Another new trend of using circularly polarized light is the study of the valley-dependent circular dichroism in two-dimensional semiconductors, recently observed in MoS$_2$ \cite{Mak2013} and WSe$_2$ \cite{Jones2013} by circularly polarized photoluminescence. 

Circularly polarized light is commonly produced (or sensed) using the combination of a linear polarizer and a $^1/_4$ wave retarder. Linear polarizers made from free-standing wire grids or deposited metal strips on a transparent substrate perform well in the far infrared.  On the other hand, $^1/_4$ wave retarders that exploit birefringent waveplates perform well only over a narrow range of frequencies \cite{Murdin2013}, limiting their usefulness to monochromatic sources such as lasers, as reported in Ref \cite{Jenkins2010}, although careful design based on multiple waveplates has been demonstrated to extend the operating frequency range \cite{Masson2006}.  An approach for making an achromatic retarder follows the principle of a Fresnel rhomb where the phase shift upon total internal reflection depends only on the refractive index and angle of incidence. However, due to the considerable optical thickness of the device, the choice of material is limited to those having a very low absorption coefficient.  In this work, we discuss the development of a Fresnel rhomb-type $^1/_4$ wave retarder using silicon and demonstrate its capabilities by performing photo-induced magneto-spectroscopy on GaAs.

\section{Development and characterization of a silicon $^1/_4$ wave retarder}
\subsection{Operating principle and design}

Materials that are both transparent and isotropic in the spectral range of interest are necessary for a retarder design. The phase shift between the $s$ and $p$ polarizations from total internal reflections is determined by the refractive index of the retarder material and the angle of incidence. An optimal design therefore involves careful choices of these two parameters. Maintaining a frequency-independent phase shift implies an almost constant refractive index in the frequency range of interest. We also desire a compact optical system that can be readily added to an existing experimental setup while introducing minimal optical loss. Based on these considerations we have chosen high-resistivity silicon (refractive index $n = 3.42$ in the far infrared) for our retarder. A double-Fresnel rhomb design [Figure 1(a)] using silicon was proposed by two of us for a far-infrared full Mueller Matrix ellipsometry \cite{Kang2011,Stanislavchuk2013}.  A consequence of the high refractive index is a rather large phase shift for a single internal reflection.  To manage this, an internal reflectance angle of 27$^{\circ}$ was chosen to yield a $5\pi/8$ phase shift.  Thus, for a design where the beam experiences four such reflections [Figure 1(b)], a total phase shift of $5\pi/2$ ($= 2\pi+ \pi/2$), i.e. a net of $^1/_4$ wave occurs, producing circularly polarized light when the incident light is linearly polarized and oriented to yield equal amplitudes of $s$ and $p$ reflections.  

\begin{figure}[t]
\includegraphics[scale=0.62]{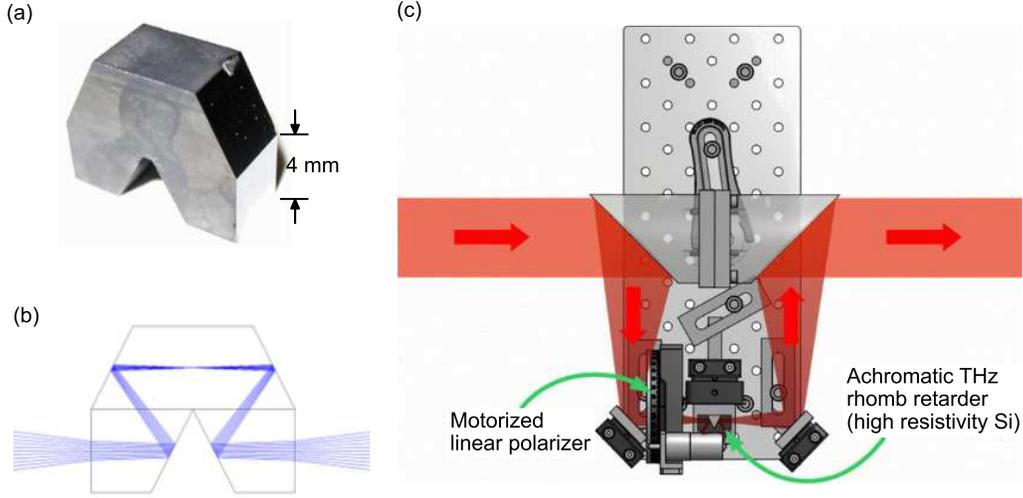}   
\caption{(a) Double-Fresnel rhomb $^1/_4$ wave retarder made of high resistivity silicon. The reflecting surfaces in the beam path are mechanically polished to reduce beam scattering. (b) Cross section of the rhomb retarder showing the designed beam path (blue lines). (c) Optical assembly for the circular polarizer. The motorized linear polarizer just upstream of the retarder sets the polarization angle of the incident linearly polarized beam to be $\pm 45^{\circ}$ with respect to the horizontal direction (defined according to the right-hand rule), yielding right or left circularly polarized light.} 
\label{FIG1}
\end{figure}

We implemented this rhomb retarder design using the optical assembly shown in Figure 1(c). A pair of 50 mm aperture off-axis paraboloidal mirrors focus a collimated incident beam through the linear polarizer and onto the retarder, and re-collimates the transmitted beam back onto the original optical path. The polarizer transmission angle is set to +45$^{\circ}$ or $-45^{\circ}$ with respect to the horizontal direction to yield equal amplitudes of $s$ and $p$ polarized reflected amplitudes.  The linear polarizer's rotation is motorized to allow switching between left or right circularly polarized light. We note that the reflections from the aluminum coated off-axis paraboloids have minimum effects on the polarization state in the far infrared. All optical components are assembled on a 100 mm $\times$ 250 mm breadboard, which can be conveniently slid into and out from the beam path. The optics are sufficiently large to work with common commercial FTIR spectrometers having 40 mm diameter optics.  

\subsection{Retarder performance}
The retarder system was tested in the far infrared for use mostly below a frequency of 200 cm$^{-1}$. Figure 2 shows the transmittance of the retarder between 20 -- 250 cm$^{-1}$, measured at room temperature.  Above $\sim$100 cm$^{-1}$ the two-phonon absorption in silicon \cite{Shirley2007} reduces significantly the system's overall transmittance. 

\begin{figure}[t]
\centering
\includegraphics[scale=0.5]{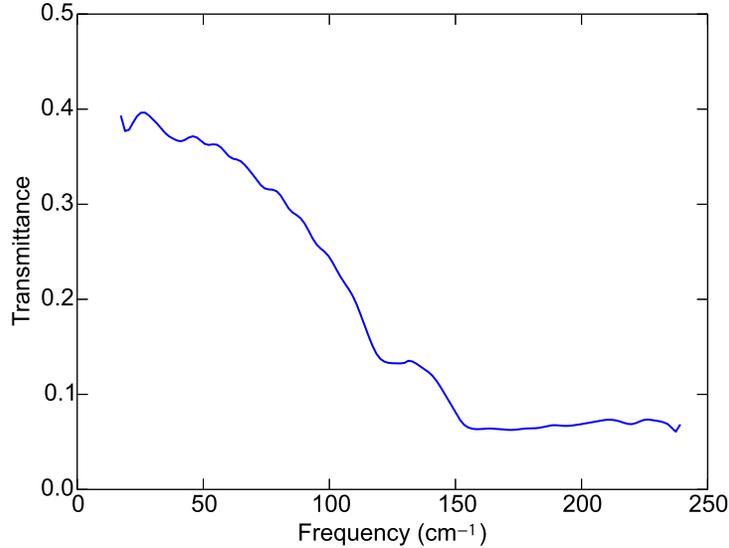}   
\caption{Transmittance of the silicon retarder at room temperature.} 
\label{FIG2}
\end{figure}

\begin{figure}[htbp]
\centering
\includegraphics[scale=0.68]{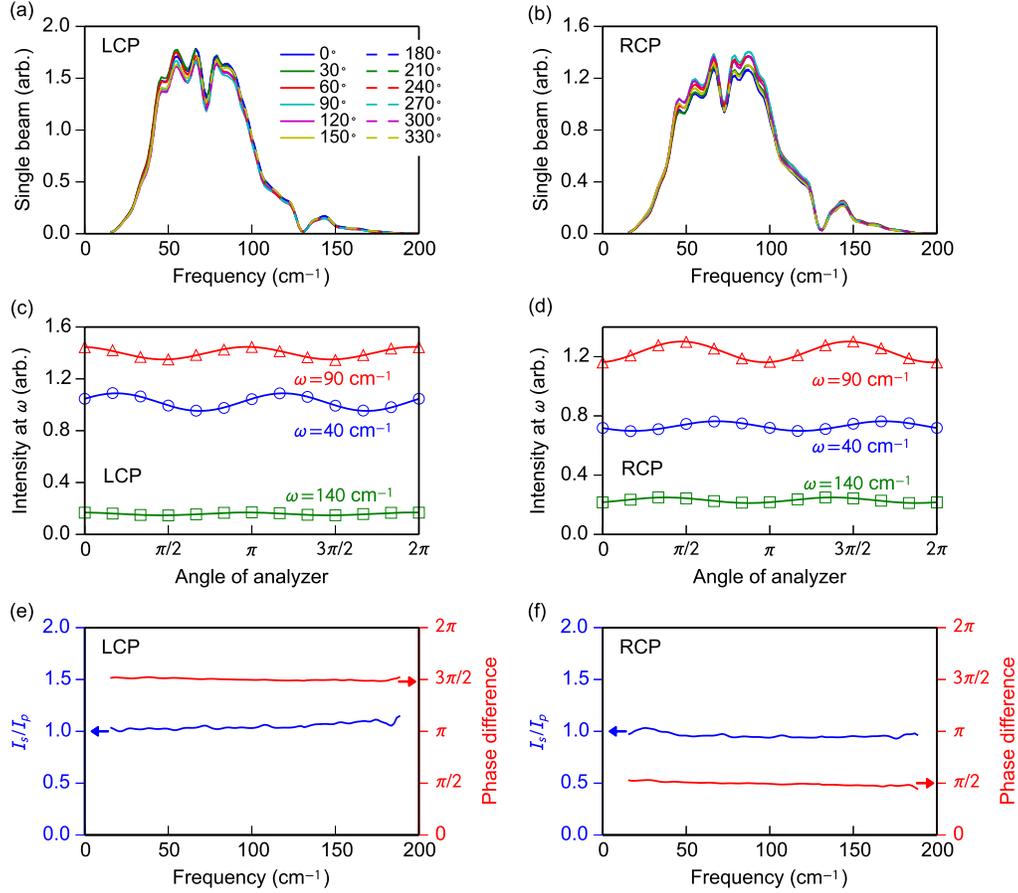}   
\caption{(a,b) Single-beam spectra of the transmitted beam through the circular polarizer and a linear polarizer (i.e. analyzer) downstream of the retarder. The analyzer angle was varied from 0$^{\circ}$ to 360$^{\circ}$ to characterize the polarization state. (c,d) Intensity at 40 cm$^{-1}$, 90 cm$^{-1}$ and 140 cm$^{-1}$ read off from the single beams as a function of the analyzer angle. The solid lines are fits using Equation~\ref{eq_iten_ang}. (e,f) Intensity ratio of, and phase difference between, the $s$ and $p$ components of the beam through the circular polarizer. LCP: left circularly polarized light. RCP: right circularly polarized light.} 
\label{FIG3}
\end{figure}

We analyzed the polarization state of the beam output from the circular polarizer (i.e. the retarder plus the upstream polarizer) using a linear polarizer (analyzer) downstream of the assembly. We define the intensity ratio of, and phase shift between, the $s$ and $p$ components of the beam output from the circular polarizer using the convention of ellipsometric angles $\Psi$ and $\Delta$,
\begin{equation}
\tan\Psi = \sqrt{\frac{I_p}{I_s}},
\end{equation}
\begin{equation}
\Delta = \delta_p - \delta_s,
\end{equation}
where $I_s$ ($I_p$) and $\delta_s$ ($\delta_p$) are the intensity and phase of the $s$ ($p$) component, respectively. We set the polarizer just upstream of the silicon retarder to an angle $P$ (either 45$^{\circ}$ or $-45^{\circ}$). We rotated the analyzer to angle $A$ and recorded the transmitted single-beam spectrum, shown in Figure 3(a,b). The detected intensity can be written via Fourier coefficients $\alpha$ and $\beta$ as \cite{Humlicek2005}
\begin{equation}
I\propto 1+\alpha\cos{(2A)}+\beta\sin{(2A)},\label{eq_iten_ang}
\end{equation}
where $\alpha$ and $\beta$ are related to the ellipsometric angles through
\begin{equation}
\tan{\Psi} = |\tan P|\sqrt{\frac{1+\alpha}{1-\alpha}},
\end{equation}
\begin{equation}
\cos{\Delta} = \mathrm{sgn}(P)\frac{\beta}{\sqrt{1-\alpha^2}}.
\end{equation}
Fitting the $A$-dependent intensity at each frequency using Equation~\ref{eq_iten_ang} [example fits shown in Figure 3 (c,d)], we extracted $I_p/I_s$ and $\delta_p - \delta_s$ over the measured frequency range, shown in Figure 3 (e,f). The intensity ratio, being very near to unity, along with the observed phase shift of $\pi/2$ and $3\pi/2$ (or $-\pi/2$), show that a high degree of circular polarization is produced.

\subsection{Example: Photo-induced magneto-spectroscopy of GaAs}
We next demonstrate the usefulness of this achromatic retarder in a study of electronic excitations in GaAs when placed in a magnetic field. A Ti:sapphire laser with photon energy tuned to just above the band gap energy creates low-energy electrons in the conduction band and holes in the valence band.  At low temperatures, some of the electrons and holes can bind to form reasonably long-lived excitons.  In a magnetic field $B$, the conduction band becomes quantized into Landau levels whose spacing is proportional to $eB/m^*$, where $e$ and $m^*$ are the electron charge and effective mass, respectively. The exciton's electronic levels are modified by an applied magnetic field in a manner analogous to the hydrogen atom.  In both cases the far-infrared absorption associated with electronic transitions are affected by the applied field.  This includes a dependence on the handedness of the incident circularly polarized light as determined by the field geometry. 

In our experiment, a bulk semi-insulating 100-oriented GaAs wafer of 400 $\mu$m thickness was cooled to 5 K in a superconducting magnet. Spectroscopy was performed in the Faraday geometry, where magnetic fields up to 10 T were applied parallel to the incident beam and perpendicular to the wafer surface.  About 200 mW of near-infrared beam (at 795 nm wavelength) from a Ti:sapphire laser was focused into a 3 mm diameter spot to photo-excite electrons from the valence band to the conduction band. Transmission through the sample was measured using a Bruker 66 spectrometer with a bolometric detector.  The far-infrared source was synchrotron radiation from Beamline U4IR of the NSLS, Brookhaven National Laboratory.  We used a conventional photo-induced measurement scheme where the sample's far-infrared transmission was measured with both the laser off and then on to produce transmission signals $ T_{\mathrm{off}}$ and $ T_{\mathrm{on}}$, respectively.  The resulting photo-induced transmission spectra, defined as $-\Delta T/T = - (T_{\mathrm{on}} - T_{\mathrm{off}})/T_{\mathrm{off}}$, reveals the increased absorption due to the laser-excited electrons and holes.  
 
\begin{figure}[htbp]
\centering
\includegraphics[scale=0.68]{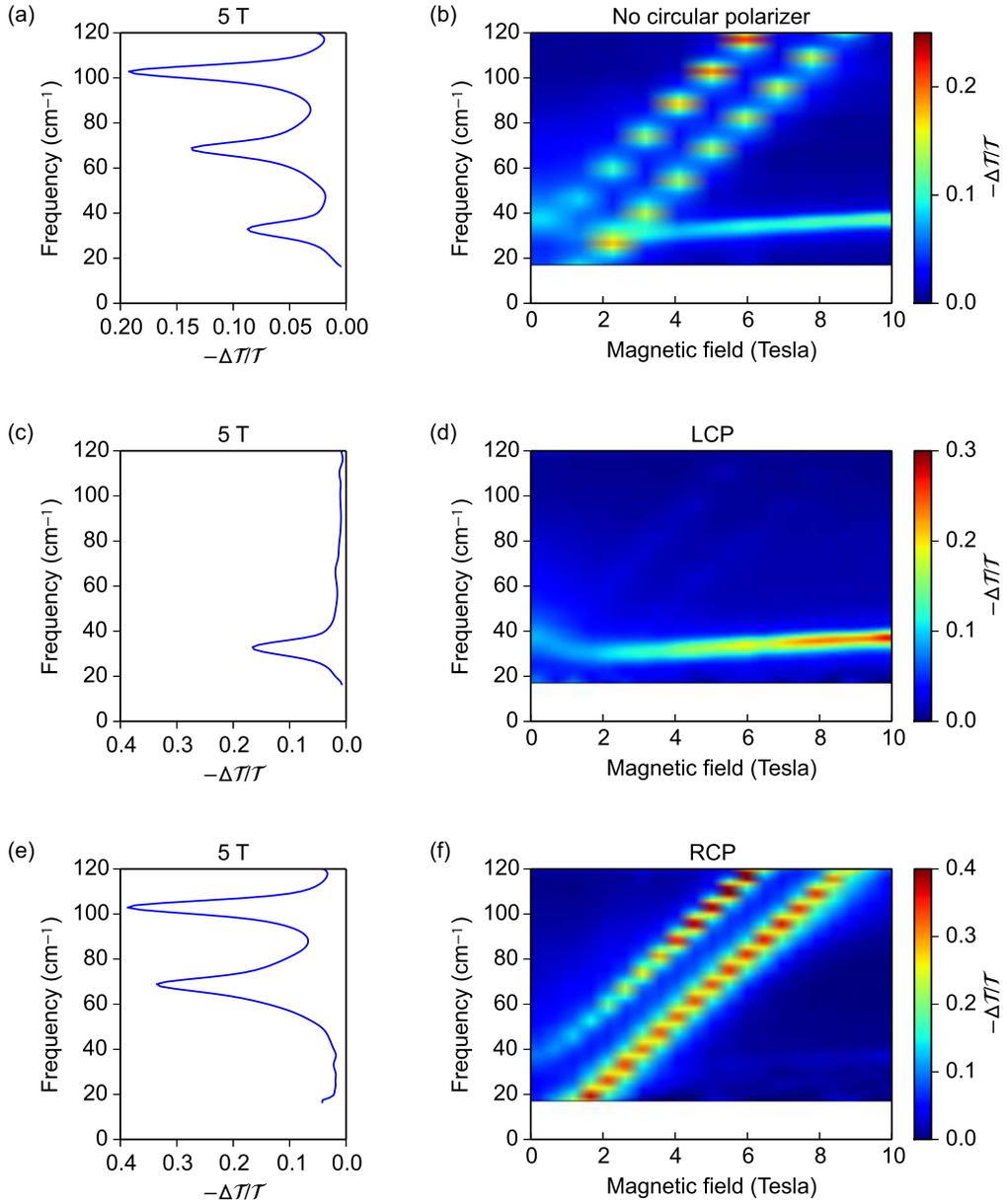}   
\caption{Photo-induced transmission $-\Delta T/T$ of GaAs in a perpendicular magnetic field of 5 T and at 5 K measured (a) without the circular polarizer, (c) using left circularly polarized light, and (e) using right circularly polarized light. (b), (d), and (f) are the corresponding field-dependent results, taken with a step of 1 T in (b) and 0.5 T in (d,f).} 
\label{FIG4}
\end{figure}
 
Sets of measurements were performed both with and without the circular polarizer for comparison. Representative spectra at $T = 5$~K and $B = 5$~T are shown in Figure 4(a), (c), and (e). The corresponding field-dependent results are shown in Figure 4(b), (d), and (f). Three strong absorption features are generally observed, all of which shift position as the field is varied.  The frequency of one feature scales directly as $eB/m^*$ with zero intercept, so can be immediately identified as the electron cyclotron resonance.  The resulting effective mass is $0.067m_0$ (where $m_0$ is the bare electron mass), in good agreement with that of conduction band electrons for GaAs.  The absorption feature near 37 cm$^{-1}$ for $B=0$ is assigned to the $1s$ to $2p$ transition of an electron bound to a hole as an exciton.  This feature splits into two branches for $B\neq 0$, each with its own dependence on $B$.  This can be understood in terms of the orbital momentum state of the $2p$ final state level where the nominally degenerate levels having quantum numbers $-1$, $0$ and $+1$ become split in the field.  The selection rules for absorbing light require a change in the angular momentum quantum number of ±1, thus the allowed transitions from the $1s$ level (0 angular momentum) are to either the $2p_{-1}$ or $2p_{+1}$. Using circularly polarized light, we were able to distinguish the split exciton absorption features: The $1s \rightarrow 2p_{+1}$ ($1s \rightarrow 2p_{-1}$) transition changes angular momentum by +1 ($-1$) and therefore is only active for right (left) circularly polarized light, as shown in Figure 4(f) [Figure 4(d)]. The cyclotron resonance involves the transitions between Landau levels and changes the angular momentum by +1; given the experimental configuration, it is active only for right circularly polarized light, as shown in Figure 4(f).  The imperfect circular polarization is a likely cause for the small absorption features observed for the opposite polarization which, ideally, should be absent. 

\section{Conclusions}
We have developed an achromatic $^1/_4$ wave retarder, based on the principle of a Fresnel rhomb, made from high resistivity silicon and for use in the far-infrared spectral range (below about 200 cm$^{-1}$).  Measurements to characterize the resulting polarization state confirmed a high degree of circular polarization. We demonstrated its performance by distinguishing the handedness of transitions involving electron cyclotron resonance and split exciton levels in GaAs in a magnetic field. This device can be useful for probing the intricate electronic states in novel condensed matter systems and possibly for studying vibrational circular dichroism in molecules associated with biological systems.

\bigskip
\noindent
\textbf{Acknowledgements}
\vspace{2mm}

This research was supported by the U.S. Department of Energy under contract DE-AC02-98CH10886 at Brookhaven National Laboratory. Technical contributions from Gary Nintzel are gratefully acknowledged.





\end{document}